%% file: Rogue.tex
\documentclass[conference]{IEEEtran}
\IEEEoverridecommandlockouts

\usepackage{cite}
\usepackage{amsmath,amssymb,amsfonts}
\usepackage{algorithmic}
\usepackage{graphicx}
\usepackage{textcomp}
\usepackage{xcolor}
\usepackage{pifont}

\usepackage{hyperref}
\usepackage{siunitx}
\usepackage{multirow}
\usepackage{booktabs,tabularx,makecell,array}

\usepackage[nohyperlinks,nolist]{acronym}
\usepackage{threeparttable}
\usepackage{subcaption}
\usepackage{float}
\usepackage{caption}
\usepackage{tikz}
\usetikzlibrary{arrows.meta}

\newcommand{\rogueurl}{\url{https://github.com/edge-group-polito/Mage-and-Rogue}}   

\begin{document}

\input{acronyms}

\title{Rogue: A Streaming CGRA With Versatile ISA for Integer Approximation of Activation Functions}

\author{
    \IEEEauthorblockN{
        Alessio Naclerio\IEEEauthorrefmark{1}\IEEEauthorrefmark{2},
        David Mallasén\IEEEauthorrefmark{1},
        Davide Schiavone\IEEEauthorrefmark{1},
        Mariagrazia Graziano\IEEEauthorrefmark{2},
        David Atienza\IEEEauthorrefmark{1}
    }

    \IEEEauthorblockA{
        \IEEEauthorrefmark{1}ESL, École Polytechnique Fédérale de Lausanne (EPFL), Lausanne, Switzerland, [name].[surname]@epfl.ch\\
        \IEEEauthorrefmark{2}DET, Politecnico di Torino, Torino, Italy, [name].[surname]@polito.it
    }
}

\maketitle

\begin{abstract}
The growing deployment of Transformer models on edge platforms has exposed non-linear activation functions as a critical performance and energy bottleneck. Although integer approximation strategies have emerged to reduce reliance on floating-point computation, existing accelerators target fixed activations or single approximation families, sacrificing the flexibility these evolving strategies demand. We propose \textbf{Rogue}, a streaming DMA-coupled CGRA implementing a dedicated ISA that captures recurring computational patterns across integer approximation families. Tight integration with the system DMA enables efficient streaming computation without additional pressure on the system bus, and the generality of the ISA extends naturally to linear kernels, letting Rogue serve both workload classes on a unified fabric. Several implementations spanning different area-performance trade-offs are evaluated on kernels covering polynomial, base-substitution and piecewise strategies. Integrated in an open-source microcontroller, Rogue delivers up to 137$\times$ speedup and 52$\times$ energy savings over a throughput-optimized RISC-V CPU baseline, and, at iso-technology, is over 2$\times$ smaller and up to 5$\times$ more area efficient than the closest state-of-the-art CGRA while covering a broader range of approximation families.
\end{abstract}

\begin{IEEEkeywords}
CGRA, Activation Functions, RISC-V, Streaming Computation, DMA
\end{IEEEkeywords}

\section{Introduction}
Compute-intensive \ac{ML} models increasingly run on resource-constrained, energy-efficient edge platforms, enabled by smaller models and heterogeneous systems offloading \ac{GeMM} to programmable accelerators. However, the shift from \acp{CNN} to attention-based Transformers has moved the bottleneck into the activation layers~\cite{belano_flexible_2025, wang_vexp_2025}. \ac{ReLU} dominated between 2015 and 2017, but its usage had dropped to 5\% by 2022~\cite{andri_flex-sfu_2025}, replaced by more complex non-linearities such as \ac{Tanh}, \ac{GELU} and Softmax. Hence, research has turned to integer approximations~\cite{kim_i-bert_2021, li_i-vit_2023, stevens_softermax_2022} and to dedicated accelerators, which typically fix one activation or one approximation strategy, sacrificing the flexibility these rapidly evolving strategies demand. Yet these families share a common set of primitives within a loop structure with minimal data dependencies, making them well suited to \ac{CGRA}-based acceleration.

We thus propose Rogue, an open-source streaming \ac{CGRA} that performs computations on the fly through the system \ac{DMA}, targeting efficient integer approximation of activation functions on \ac{MCU}-class platforms. Unlike recent \acp{CGRA} targeting integer activations~\cite{prasad_nx-cgra_2025, yin_s_2026, qin_picachu_2025}, Rogue spans multiple approximation families and extends naturally to \ac{GeMM} computation, enabling end-to-end inference acceleration on a single fabric. The main contributions of this paper are:

\begin{itemize}
    \item We introduce a versatile \ac{ISA} capturing recurring computational patterns across the main families of integer activation approximations (polynomial, base-substitution, and piecewise polynomial), while remaining extensible to emerging strategies.

    \item We present Rogue, an open-source streaming \ac{CGRA} tightly integrated with the \ac{DMA} of an \ac{MCU} platform, reusing a bandwidth envelope sufficient to sustain peak throughput without extra bus pressure or a dedicated memory interface. Rogue is published as open-source at \rogueurl.

    \item We explore the design space across multiple Rogue implementations, reaching up to 137$\times$ and 52$\times$ performance and energy gains over the \textit{cv32e40p}~\cite{gautschi_near-threshold_2017} RISC\nobreakdashes-V CPU on integer activation kernels and, combining \ac{GeMM} and activation support, 7.1$\times$ and 5.3$\times$ on a full Transformer encoder layer.

    \item We compare Rogue with state-of-the-art activation accelerators, showing that it is over 2$\times$ smaller and up to 5$\times$ more area efficient than the closest \ac{CGRA} at iso-technology, while also covering a broader set of strategies.
\end{itemize}

\section{Related Work}
Four main approximation strategies have been adopted in activation accelerators. \textbf{\acp{LUT}}~\cite{xie_twofold_2020} store pre-computed values directly, typically needing substantial memory for high accuracy. \textbf{\ac{PA}}~\cite{kim_i-bert_2021, lin_towards_2021, qin_picachu_2025} uses polynomials realized as multiply-add sequences. High orders become area-hungry, but i-BERT~\cite{kim_i-bert_2021} attains high accuracy with second-order polynomials for integer \ac{GELU} and Softmax. \textbf{\ac{PWPA}}~\cite{li_low-cost_2020, prasad_pace-lite_2025, andri_flex-sfu_2025, du_algorithm-hardware_2025} and \textbf{\ac{PWLA}}~\cite{baccelli_nacu_2020, yu_nn-lut_2022} store few coefficients for low-order polynomials and linear segments, balancing memory against compute, with coefficients and breakpoints usually configurable to retarget functions. Lastly, several works~\cite{li_i-vit_2023, hu_i-llm_2024, islamoglu_ita_2023} apply \textbf{base-substitution} from $e$ to $2$ to exponential-based activations, turning much of the computation into bit shifts.

Accelerators also differ in coverage, most targeting one function or a narrow set. Early \ac{LSTM} works~\cite{zamanlooy_efficient_2014, leboeuf_high_2008} target fixed-point \ac{Tanh} and Sigmoid. Among Transformer accelerators, SoftAct~\cite{fu_softact_2024} specializes in integer Softmax and SoftEx~\cite{belano_softex_2025} in \ac{FP} Softmax, later extended to \ac{GELU}~\cite{belano_flexible_2025}, whereas ITA~\cite{islamoglu_ita_2023} embeds a Softmax unit inside a \ac{GeMM} accelerator. Specialization is efficient, but recent work stresses adaptability to a broader, evolving function set. NACU~\cite{baccelli_nacu_2020} reuses one \ac{PWLA} exponential datapath for Sigmoid, \ac{Tanh} and Softmax, NN-LUT~\cite{yu_nn-lut_2022} generalizes \ac{PWLA} with data-driven breakpoint placement, and Flex-SFU~\cite{andri_flex-sfu_2025}, PACE-Lite~\cite{prasad_pace-lite_2025} and the works in~\cite{du_algorithm-hardware_2025, li_low-cost_2020} employ \ac{PWPA} segments up to quadratic order across many activations.

Several recent works propose \acp{CGRA} as reconfigurable fabrics for activations. PICACHU~\cite{qin_picachu_2025} integrates \ac{FP} and \ac{INT} operators in its \acp{PE} for the multiply-add and reduction patterns common in \ac{PA}, but delegates \ac{GeMM} to a separate systolic array. NX-CGRA~\cite{prasad_nx-cgra_2025} and STELLA~\cite{yin_s_2026} instead unify \ac{INT} \ac{GeMM} and activation approximation on the same fabric, while relying on \ac{PA} alone.

In contrast, Rogue's \ac{ISA} captures the patterns shared by all these families, spanning \ac{PA}, \ac{PWLA}, \ac{PWPA} and base-substitution on a lightweight streaming fabric that extends naturally to \ac{GeMM}.

\section{Proposed ISA}
Different integer approximations share a common set of computational primitives. \ac{PA} relies on chained multiply-add sequences to combine polynomial terms with their coefficients. \ac{PWLA} and \ac{PWPA} add a segment-selection step that compares the input against breakpoints to identify the active interval and fetch its coefficients. All three need shifts for rescaling, divisions for normalization and clipping to valid ranges, and base-substitution converges on the same operator set, with shifting replacing exponentiation. Hence, a dedicated \ac{ISA} can expose these recurring patterns. From the algorithmic structure of the four families, we identified operator groups that (i) form the computational core of a family and/or recur across at least two, and (ii) fuse into one multi-operation instruction, collapsing two \ac{DFG} nodes into one. Although implementation-independent by definition, the \ac{ISA} suits \acp{CGRA} particularly well, as each fused pair saves one operation and one \ac{PE} per kernel. Fusing operators while limiting variable operands improves \ac{PE} utilization and efficiency~\cite{del_sozzo_exploration_2024} and reduces interconnection complexity.

The proposed \ac{ISA} employs 32-bit instructions packing a 6-bit opcode, two 4-bit source-operand selectors, a 2-bit packed-\ac{SIMD} mode field, and a 16-bit constant operand. \texttt{S-type} instructions take one or two variable operands, whereas \texttt{C-type} ones add a third, fixed to the instruction's constant $C$. Since $C$ travels with the instruction rather than through the interconnect, \texttt{C-type} instructions widen the fusable operator set while leaving the variable-operand count, and hence the interconnection complexity, unchanged. Table~\ref{tab:isa} lists the instructions with their latencies in Rogue, which the \ac{ISA} does not constrain. Beyond standard integer arithmetic and shifts, each fused instruction maps onto one of the recurring patterns above: squared-term computation in exponential approximations (\texttt{ADDPOW}, \texttt{SUBPOW}); sign handling for symmetric functions such as \ac{Tanh} and Sigmoid, on the input (\texttt{SGNSEL}) and output (\texttt{SGNCSUB}) sides; normalization (\texttt{ABSDIV}, \texttt{ABSREM}, \texttt{CADDDIV}); Horner polynomial evaluation and multiply-with-offset for \ac{PA} (\texttt{CMULADD}, \texttt{CADDMUL}, \texttt{ADDCMUL}), extended to \ac{PWLA} and \ac{PWPA} segments by \texttt{ADDCMULPW}, which sources both coefficients from the coefficient \acp{RF}; and post-multiplication rescaling (\texttt{MULCARSH}, \texttt{CLSHSUB}). Finally, \texttt{BPSEL} supports the binary-tree interval search of \ac{PWLA} and \ac{PWPA}, comparing the input against a breakpoint from a dedicated \ac{RF} and returning the index with an appended comparison bit.

\begin{table}[htbp]
    \centering
    \begin{threeparttable}
        \caption{Proposed \ac{ISA} instructions and their latency in Rogue. The \textit{PE} column reports which \ac{PE} implements each instruction, \textit{all} marking the subset available in every \ac{PE} and the other entries requiring the corresponding extension.}
        \label{tab:isa}
        \begin{tabular}{@{}ccc
        !{\vrule\vrule}
        cc@{}}
            \toprule
            Type & Name & Function & PE & Latency\\
            \midrule
            \multirow{11}{*}{\texttt{S}} & \texttt{ABS} & $abs(a)$ & all & 1\\
             & \texttt{ADD}, \texttt{SUB}, \texttt{MUL}    & $a [\mathrel{+, -, *}] b$ & all & 1\\
             & \texttt{LSH}, \texttt{ARSH}, \texttt{LRSH}\tnote{a}  & $a [\mathrel{<<, >>}] b$ & all & 1\\
             & \texttt{MAX}, \texttt{MIN}  & $max(a,b), min(a,b)$ & all & 1\\
             & \texttt{DIV}, \texttt{REM}  & $a [\mathrel{/, \%}] b$ & Div & $L_{div}$\tnote{b}\\
             & \texttt{SGNSEL} & $a > 0 \mathrel{?} a \mathrel{:} b$ & all & 1 \\
             & \texttt{ABSDIV} & $abs(a) \mathrel{/} b$ & Div & $L_{div}+1$\tnote{b}\\
             & \texttt{ABSMIN} & $abs(a) > b \mathrel{?} b \mathrel{:} a$ & all & 1 \\
             & \texttt{ABSREM} & $abs(a) \mathrel{\%} b$ & Div & $L_{div}+1$\tnote{b}\\
             & \texttt{ADDPOW} & $(a + b) * (a + b)$ & all & 2 \\
             & \texttt{SUBPOW} & $(a - b) * (a - b)$ & all & 2 \\
             & \texttt{BPSEL} & $a \leq \text{BP[b]} \mathrel{?} (b, 0) \mathrel{:} (b, 1)$\tnote{c} & BP & 1 \\
            \midrule
            \multirow{7}{*}{\texttt{C}} & \texttt{SGNCSUB}  & $a > 0 \mathrel{?} (C - b) \mathrel{:} (b - C)$ & all & 1 \\
             & \texttt{CADDMUL}  & $(a + C) * b$ & all & 2 \\
             & \texttt{ADDCMUL}  & $(a + b) * C$ & all & 2 \\
             & \texttt{ADDCMULPW}  & $(a + CF1[b]) * CF2[b]$\tnote{c} & Cf & 2 \\
             & \texttt{CMULADD}  & $(C * a) + b$ & all & 2 \\
             & \texttt{CADDDIV}  & $(a + C) \mathrel{/} b$ & Div & $L_{div}+1$\tnote{b}\\
             & \texttt{MULCARSH} & $(a * b) \mathrel{>>>} C$ & all & 2 \\
             & \texttt{CLSHSUB}  & $(a \mathrel{<<} C) - b$ & all & 2 \\
            \bottomrule
        \end{tabular}
        \begin{tablenotes}
            \footnotesize
            \item[a] \texttt{ARSH} uses signed right shifting, while \texttt{LRSH} unsigned.
            \item[b] Divider latency $L_{div}=\frac{32}{\log_2(radix)}$ is the same for serial and pipelined dividers, but pipelined ones have higher throughput.
            \item[c] \textit{BP} is the breakpoints \ac{RF}. \textit{CF1} and \textit{CF2} are \acp{RF} for polynomial coefficients.
        \end{tablenotes}
    \end{threeparttable}
\end{table}

\begin{figure*}[t]
    \centering
    \includegraphics[width=\linewidth]{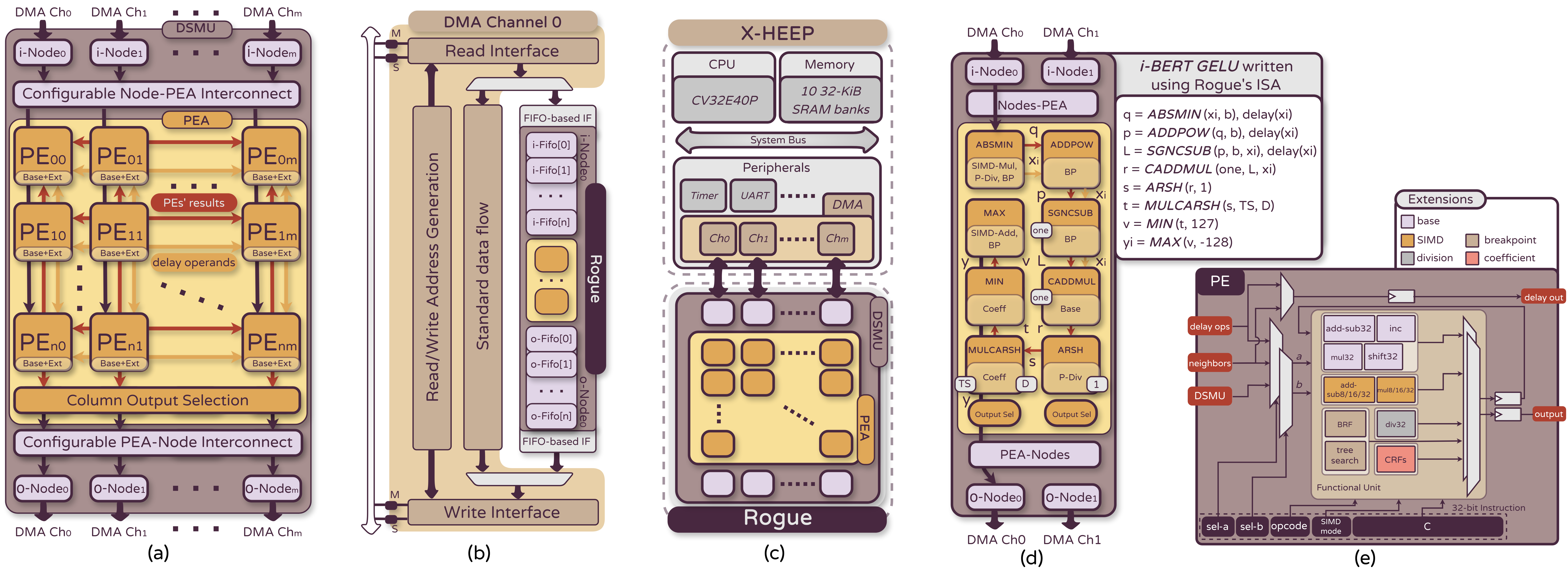}
    \caption{(a) Rogue's design, (b) its integration within a DMA channel and (c) in the X-HEEP \ac{MCU}, (d) \ac{GELU} described with the proposed \ac{ISA} and mapped onto the \textit{P4x2} variant, and (e) the base \ac{PE} datapath with its extensions.}
    \label{fig:rogue}
\end{figure*}

\section{Rogue Architecture}
\label{sec:arch}
Rogue is a streaming \ac{CGRA} designed for integer approximation of activation functions, tightly coupled to existing system \acp{DMA}~\cite{terzano_just_2025, benz_high-performance_2024} through a \ac{FIFO}-based interface. As shown in Figure~\ref{fig:rogue}(a), its architecture revolves around two key modules: the \ac{PEA} and the \ac{DSMU}.

\subsection{Data Stream Management Unit}

The \ac{DSMU} holds a configurable set of \ac{I/O} nodes and manages their interconnection with the external \acp{DMA} and the \ac{PEA}. As Figure~\ref{fig:rogue}(b) shows, it links Rogue to the \ac{DMA} channels so that data read by the \ac{DMA} is processed by the \ac{PEA} and written to the transfer destination. Each \ac{I/O} node is \ac{FIFO}-based and tied to a specific \ac{DMA} channel, yet the \ac{DSMU} supports direct and multiplexed \ac{FIFO}-to-column and column-to-\ac{FIFO} paths, and synchronizes interdependent input streams so that \acp{PE} receive correctly ordered data.

\subsection{Processing Element Array}
The \ac{PEA} is Rogue's core computational module. It implements the proposed \ac{ISA} and supports streaming, elastic execution and internal handling of intra-iteration dependencies.

\subsubsection{\ac{PEA} Implementation}

Rogue's \ac{PEA} is a 2-D array of nearest-neighbor-connected \acp{PE}, without crossbars or switchboxes, configurable in array size, \ac{PE} type and inter-\ac{PE} connectivity, all types sharing a base design extensible with specialized capabilities for different area-performance trade-offs. The base \ac{PE} implements every \texttt{S} and \texttt{C} instruction except the division-based and breakpoint/coefficient-selection ones, drawing operands $a$ and $b$ from neighboring \acp{PE}, the column input of a \ac{DSMU} node, or \textit{delay operands} forwarded by neighbors. These values form a delay chain resolving complex intra-iteration dependencies without breaking the stream. The base \ac{PE} holds a 32-bit adder/subtractor, a 32-bit multiplier, a 32-bit arithmetic right barrel shifter and an incrementer. Comparisons reuse the adder carry-out, and left shifts reuse the right shifter through bit reversal. Every base \ac{PE} instruction completes in one or two cycles. Since multi-operation instructions combine two operators, they are pipelined within the \ac{PE}, the intermediate result being registered and consumed in the next cycle so a new operand still enters every cycle at an \ac{II} of one.

Three families of extensions enhance the base \ac{PE}. A \textit{division} extension integrates a serial or pipelined divider, enabling division-based instructions and exposing an area-performance trade-off, as serial dividers may suffice where throughput requirements are relaxed. Serial dividers are radix-configurable, resolving $\log_2(radix)$ result bits per cycle, while pipelined ones use 32 subtractors to sustain 1~\ac{OpC} once filled, both incurring a latency of $\frac{32}{\log_2(radix)}$ cycles. A \textit{\ac{SIMD}} extension replaces the base adder or multiplier with a partitioned unit that retains full-width operation, enabling packed \ac{SIMD} variants of all addition- and multiplication-based instructions, mainly for \ac{GeMM}. Finally, \textit{breakpoint} and \textit{coefficient} extensions provide \ac{PWLA} and \ac{PWPA} support. The former embeds a breakpoint \ac{RF} (BRF) enabling the \texttt{BPSEL}-driven binary-tree interval search, the latter coefficient \acp{RF} (CRF) enabling \texttt{ADDCMULPW} for direct segment evaluation. BRF and CRF sizes scale with tree depth and polynomial order, and these \acp{PE} must form a contiguous chain preserving the dataflow from segment selection to coefficient evaluation. Extensions are independently composable, their heterogeneous combination balancing flexibility and area efficiency, as in prior \ac{CGRA} works~\cite{del_sozzo_exploration_2024, bandara_revamp_2022}. Figure~\ref{fig:rogue}(e) details the base \ac{PE} datapath and its extensions.

\subsubsection{\ac{PEA} Execution Model}
To execute efficiently at low area cost, Rogue's \ac{PEA} adopts a static-dataflow elastic model, avoiding the context-switching and configuration-memory overhead of modulo-scheduled \acp{CGRA}. Each column takes its input from a \ac{DSMU} node and data propagates through an elastic pipeline whose handshake between neighboring \acp{PE} absorbs multi-cycle stalls such as serial division. The same protocol lets the \ac{DSMU} back-pressure the input \acp{FIFO}. Each \ac{PE} selects the delay operand it forwards among the values already in the delay chain and the neighbor outputs it receives. \acp{PE} left unused are clock-gated.

\subsection{Mapping and Configuration}
\label{sec:mapping}
To be mapped onto Rogue, a kernel is expressed as a \ac{DFG} over the \ac{ISA} of Table~\ref{tab:isa} and bound to the \ac{PEA}, one \ac{PE} per node, under three constraints: (i) a node using a division, \texttt{BPSEL} or \texttt{ADDCMULPW} instruction must land on a \ac{PE} carrying the matching extension; (ii) connected nodes must land on physically adjacent \acp{PE}, with delay operands routed through each \ac{PE}'s delay chain; and (iii) each column may consume one \ac{DSMU} stream. Figure~\ref{fig:rogue}(d) shows i-BERT \ac{GELU} mapped onto a 4$\times$2 array under these constraints. This process is currently manual, and practical for kernels of 4 to 8 instructions.

Mapping yields one configuration word per \ac{PE}, plus the delay-chain selections and, where present, BRF and CRF contents, all memory-mapped and written by the host CPU over the peripheral bus. The CPU then starts a \ac{DMA} transfer routed through Rogue. Execution is data-driven, triggered and terminated by the transfer itself, with the CPU notified by the ordinary \ac{DMA} completion interrupt. No instruction is ever issued to Rogue at runtime. A \ac{DFG} exceeding the array capacity is split into sub-graphs run as successive, separately configured transfers with intermediate results staged in memory, as for the multi-transfer kernels of Section~\ref{sec:results}.

\subsection{System Integration}
Although Rogue's \ac{FIFO}-based interface suits any \ac{DMA}, this work couples it to that of the X\nobreakdash-HEEP~\cite{schiavone_x-heep_2023} \ac{MCU} (Figure~\ref{fig:rogue}(c)), an open-source 32-bit RISC\nobreakdashes-V platform with configurable memory banks, \ac{DMA} channels, CPU~\cite{gautschi_near-threshold_2017, davide_schiavone_slow_2017} and peripherals. Its system \ac{DMA} sets the maximum bandwidth through the number of channels and master ports~\cite{terzano_just_2025}, and exposes a \ac{DMA}-extension interface for inserting computational units into the data stream, which Rogue uses. Reusing the \ac{DMA}'s address generation and transfer management spares Rogue bus pressure and a dedicated memory interface, but bounds its bandwidth to the \ac{DMA}'s. This is not a bottleneck in practice. The static dataflow model spreads kernels spatially across \acp{PE}, and these kernels need only 4 to 8 \ac{ISA} instructions, a compute-to-transfer ratio low enough that a 4$\times$2 \ac{PEA} already saturates two \ac{DMA} channels, a common system \ac{DMA} size.

\begin{figure*}[ht]
    \centering
    \includegraphics[width=0.9\linewidth]{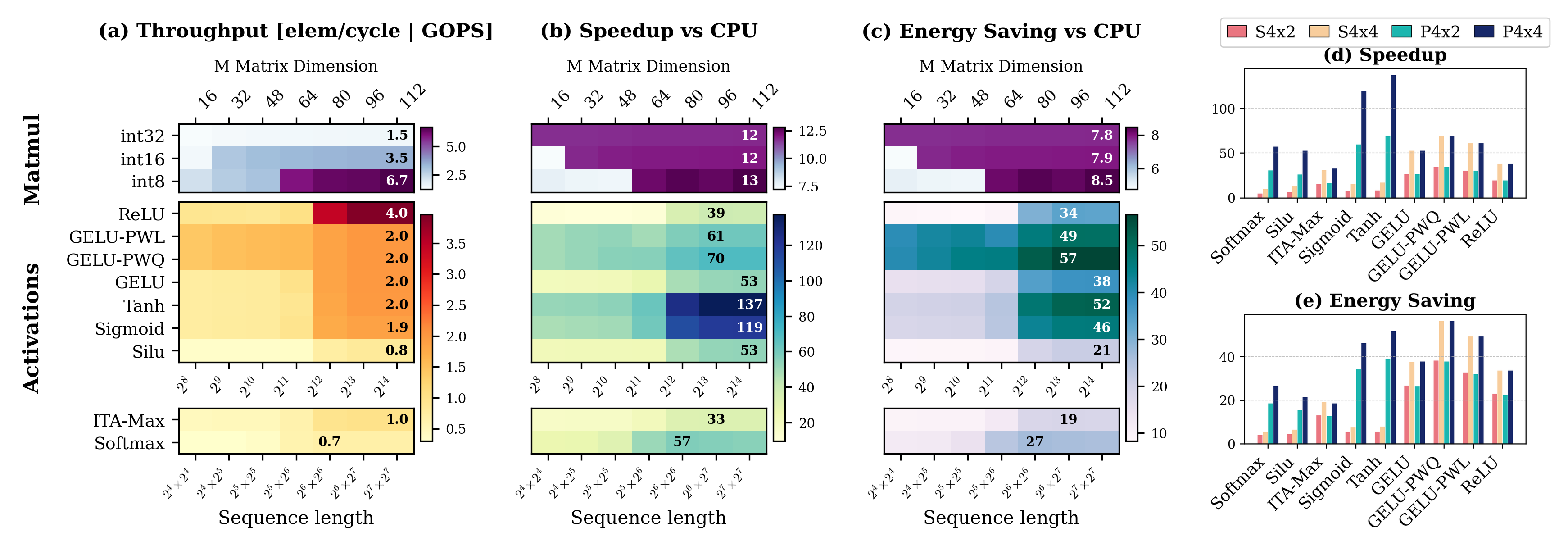}
    \caption{Rogue's throughput (a) in elem/cycle for activations and GOPS for \ac{matmul}, speedup (b) and energy savings (c) over software execution for \textit{P4x4}. \textit{Matmul} sweeps $M$ in $C[N,K]=A[N,M] \times B[M,K]$ with $N=K=32$. Softmax and ITA-Max are evaluated across varying $N_{rows} \times N_{columns}$ of the attention score matrix. The other activations are element-wise and are swept over the sequence length. Panels (d) and (e) compare speedup and energy savings across all four variants at their peak values.}
    \label{fig:rogue_p4x4_heatmaps}
\end{figure*}

\section{Experimental Setup}
\label{sec:setup}
Four Rogue variants are implemented in X\nobreakdash-HEEP-based \ac{MCU} platforms. Two adopt a 4$\times$2 \ac{PEA}, the smallest array fitting every evaluated activation. \textit{S4x2} uses two \textit{serial-division} \acp{PE} and \textit{P4x2} two \textit{pipelined-division} ones, exposing the area-performance trade-off on division-based activations. Division \acp{PE} sit at opposite corners so every division-based activation fits. One \ac{SIMD}-adder and one \ac{SIMD}-multiplier \ac{PE} serve \ac{GeMM}, whose two operands stream from separate \ac{DMA} channels into a single column. Four breakpoint and two coefficient \acp{PE} form the chain supporting \ac{PWLA} and up to quadratic-order \ac{PWPA} (PWQA), the four tree levels resolving 16 segments. Extensions can be combined on the same \ac{PE}, as Figure~\ref{fig:rogue}(d) shows for \textit{P4x2}. \textit{S4x4} and \textit{P4x4} replicate this rationale on a 4$\times$4 \ac{PEA}. The \ac{DMA} channel count matches the column count (two and four), enough for peak bandwidth.

The X\nobreakdash-HEEP systems (Figure~\ref{fig:rogue}(c)) use 320 KiB of on-chip memory, the OpenHW \textit{cv32e40p}~\cite{gautschi_near-threshold_2017} RISC\nobreakdashes-V core with the \textit{XCVPULP} ISA, and a peripheral subsystem including the \ac{DMA}. All systems, including the CPU baseline, are synthesized with Synopsys Design Compiler in a commercial 16 \si{\nano\meter} node at 330 \si{\mega\hertz} under worst-case conditions, so all comparisons are at iso-frequency and iso-technology. Cycle counts come from post-synthesis simulation, including \ac{DMA} programming and Rogue configuration, and energy is extracted with Synopsys PrimeTime from the resulting switching activity, covering \ac{DMA}, CPU, memory, bus and Rogue.

Nine integer activation kernels are evaluated: \ac{ReLU}, i-BERT's \ac{GELU} and Softmax (hereafter \ac{GELU} and Softmax), ITA-Max~\cite{islamoglu_ita_2023}, i-BERT-based \ac{Tanh}, Sigmoid and \ac{SiLU}, and \ac{PWLA}- and second-order \ac{PWPA}-based \ac{GELU}. All use 8-bit inputs with intermediate computation up to 32 bits. Rogue is also evaluated on \ac{matmul} with int8, int16 and int32 inputs. Throughput and energy are measured against the \textit{cv32e40p} running compiler-generated code targeting \textit{XCVPULP}~\cite{gautschi_near-threshold_2017}. For \ac{matmul} the baseline is hand-optimized, invoking \textit{XCVPULP} instructions directly for \ac{SIMD} execution. A Transformer encoder layer with model dimension 64, 4 heads, hidden dimension 256, and sequence length 128 is also evaluated. These dimensions match deployed keyword-spotting encoders~\cite{berg_keyword_2021} and exceed the tiny encoders targeted by \ac{MCU} deployment work~\cite{jung_optimizing_2025}. For \textit{P4x2} and \textit{P4x4}, memory is expanded to 512 KiB to fit the encoder operands. Two scenarios are considered, Rogue executing \ac{GeMM} only and \ac{GeMM} combined with Softmax and \ac{GELU}, isolating the benefit of activation acceleration. The baseline uses the Transformer-optimized kernels of~\cite{jung_optimizing_2025}.

\section{Results and Discussion}
\label{sec:results}

\input{tab/rogue_area}

\subsection{Area Evaluation}

Table~\ref{tab:area} breaks down the area of the four Rogue implementations and their X\nobreakdash-HEEP systems, the lower half giving each \ac{PE} extension's cost. Division extensions dominate, the pipelined-division \ac{PE} being 3.3$\times$ larger than the base \ac{PE} and the serial-division one 1.5$\times$. \ac{SIMD}, breakpoint and coefficient extensions instead stay within 1.15$\times$ the base area, so packed \ac{SIMD} for \ac{GeMM} and \ac{PWLA}/\ac{PWPA} support come at limited cost. \textit{P} variants are about 1.3$\times$ larger than their \textit{S} counterparts in both array sizes, buying performance on division-based activations, while 4$\times$4 variants roughly double resources and enable higher kernel parallelization. The \ac{DSMU} contributes below 5\% of Rogue's area in every variant, whereas the configuration registers are the second-largest contributor at 18-24\%. At the system level, Rogue adds from 4.4\% (\textit{S4x2}) to 11.5\% (\textit{P4x4}) of area, confirming that tight \ac{DMA} integration keeps the area cost contained.

\subsection{Performance and Energy Evaluation}
Figure~\ref{fig:rogue_p4x4_heatmaps} reports throughput, speedup and energy savings over software execution for \textit{P4x4}. Kernels differ in \ac{DMA} transfer count: \ac{ReLU}, \ac{GELU}, \ac{Tanh}, Sigmoid, \ac{GELU}-PWLA and \ac{GELU}-PWQA need one, \ac{SiLU} and ITA-Max two, and Softmax three. Single-transfer kernels reach ideal throughput at the maximum unroll factor $UF$, with $\text{OpC} = UF$, while multi-transfer ones pay reconfiguration overhead, saturating at $\text{OpC} = \frac{UF}{N_{\text{transfer}}}$. Speedup follows this trend, single-transfer kernels staying ahead, though multi-transfer ones still reach 57$\times$, 53$\times$ and 33$\times$ for Softmax, \ac{SiLU} and ITA-Max. Moreover, heavier approximations benefit most. \ac{Tanh} and Sigmoid peak at 137$\times$ and 119$\times$, also owing to the pipelined divider's advantage over the iterative count-leading-zeros one of the \textit{cv32e40p}, while division-free \ac{GELU}, \ac{GELU}-PWLA and \ac{GELU}-PWQA reach 53$\times$, 61$\times$ and 70$\times$.

\input{tab/comparison_sota}

Energy savings follow the same trend, but division-free and division-based single-transfer kernels reach comparable savings, with \ac{GELU}-PWQA and \ac{Tanh} peaking at 57$\times$ and 52$\times$, as the lower power envelope of base and coefficient \acp{PE} offsets the higher speedup of pipelined-division ones. Multi-transfer savings stay substantial, with Softmax at 27$\times$. Both metrics grow with sequence length before plateauing, as configuration overhead becomes negligible at realistic lengths.

For \ac{matmul}, Rogue \textit{P4x4} throughput scales with precision as packed \ac{SIMD} implies, up to 6.7 GOPS for int8. Since the \textit{cv32e40p} baseline also exploits packed \ac{SIMD} through \textit{XCVPULP} built-ins, that scaling does not carry into speedup and energy savings. Rogue nonetheless reaches 12-13$\times$ and 7.8-8.5$\times$, showing that \ac{DMA}-coupled streaming \ac{GeMM} acceleration stays effective against an optimized baseline.


Figure~\ref{fig:rogue_p4x4_heatmaps}(d) and (e) compare the four variants at their peak values, isolating two complementary paths to performance. Array size is the uniform one: 4$\times$4 outperforms 4$\times$2 on every metric thanks to the larger unroll factors. Pipelined division is the selective one: it pays off only on division-heavy kernels (\ac{Tanh}, Sigmoid, Softmax, \ac{SiLU}), whereas \textit{S} and \textit{P} variants match in throughput on division-free kernels. On ITA-Max, whose approximation needs far fewer divisions than i-BERT Softmax, \textit{S} variants even surpass \textit{P} ones in energy savings.

\begin{figure}[htbp]
    \centering
    \includegraphics[width=0.8\linewidth]{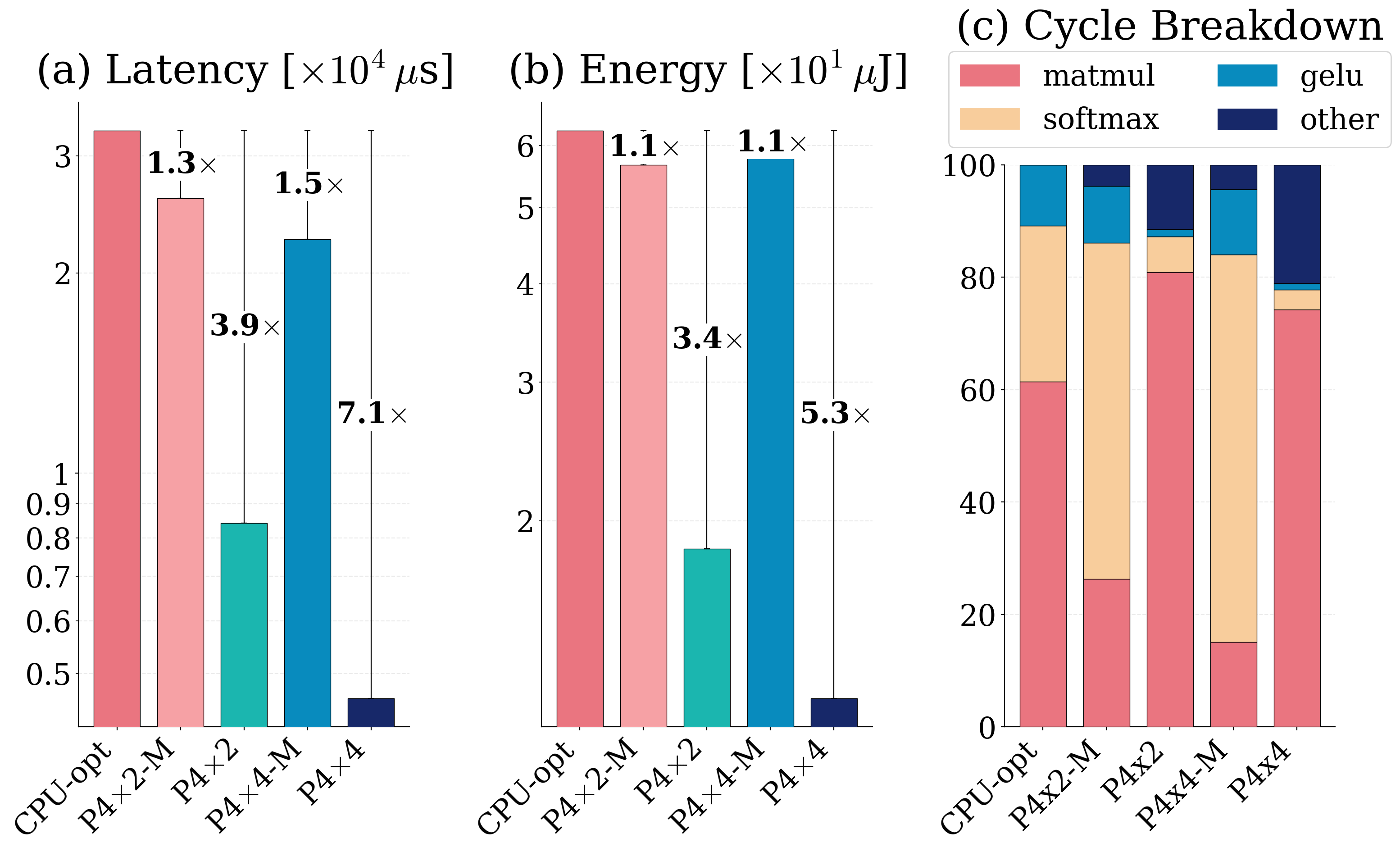}
    \caption{Latency, energy and execution time breakdown for the Transformer encoder layer.}
    \label{fig:rogue_encoder}
\end{figure}

Figure~\ref{fig:rogue_encoder} reports the latency, energy and execution-time breakdown for a Transformer encoder layer, comparing \textit{P4x2} and \textit{P4x4} against the highly optimized baseline of~\cite{jung_optimizing_2025}. The cycle breakdown in Figure~\ref{fig:rogue_encoder}(c) shows that, if only \ac{GeMM} is offloaded (\textit{P4x2-M}, \textit{P4x4-M}), Softmax and \ac{GELU} dominate the execution time, precisely the case for a unified fabric. \ac{GeMM}-only acceleration with \textit{P4x4} yields 1.5$\times$ speedup and 1.1$\times$ energy savings, whereas also accelerating Softmax and \ac{GELU} on the same fabric delivers 7.1$\times$ and 5.3$\times$.

\section{Comparison with Other Accelerators}
Table~\ref{tab:comparison-sota} compares \textit{P4x4}, the highest-performance variant, with state-of-the-art activation accelerators, throughput expressed as GEval/s (giga output samples per second). Two classes emerge, distinguished by the \textit{Linear} column. NACU~\cite{baccelli_nacu_2020}, Flex-SFU~\cite{andri_flex-sfu_2025}, PACE-Lite~\cite{prasad_pace-lite_2025} and PICACHU~\cite{qin_picachu_2025} accelerate activations alone, so they cannot run the linear operations of a layer, whereas ITA~\cite{islamoglu_ita_2023}, NX-CGRA~\cite{prasad_nx-cgra_2025} and Rogue cover a whole layer. Among the latter, Rogue is over 2$\times$ smaller at iso-technology~\cite{stillmaker_scaling_2017} and covers the widest activation set.

NX-CGRA is the closest competitor. Rogue is faster on \ac{GELU} and \ac{GeMM} and, at iso-technology, 2.5$\times$ and 5.0$\times$ more area efficient, matching its energy efficiency on \ac{GELU} and exceeding it by 1.9$\times$ on \ac{GeMM}. NX-CGRA leads only on Softmax, yet its GOPS count individual kernel operations whereas Rogue's GEval/s count complete evaluations, a mismatch penalizing Rogue in proportion to the operations per evaluation, highest for Softmax. The comparison is also the most direct, as both designs are \ac{INT}-only on closely related X\nobreakdash-HEEP-based platforms: Rogue reaches higher efficiency at lower area by replacing dedicated \acp{LSU} with the system \ac{DMA}, while its \ac{ISA} covers a wider set of approximations.

PICACHU targets the same non-linear operators on a \ac{CGRA}, but keeps \ac{GeMM} on a separate systolic array. Its 16-tile \ac{CGRA} alone occupies 190$\times10^{3}$ \si{\micro\meter\squared} scaled to 16 \si{\nano\meter}, 3.2$\times$ \textit{P4x4} at equal array size, and draws \SI{64.2}{\milli\watt} at its native 45 \si{\nano\meter} and \SI{1}{\giga\hertz}, against 1.0-1.6 \si{\milli\watt} for \textit{P4x4} at 16 \si{\nano\meter} and \SI{330}{\mega\hertz}. PICACHU therefore confirms that \acp{CGRA} suit non-linear operators, but spends each tile on activations alone, whereas Rogue keeps its \acp{PE} light enough for both activations and \ac{GeMM} within an \ac{MCU} budget.

These activation-only designs describe only part of a layer and become comparable to Rogue's once a \ac{GeMM} engine is added. That composition is costly. ITA, which combines an ITA-Max unit with a separate matmul core, occupies 131$\times10^{3}$ \si{\micro\meter\squared} at 16 \si{\nano\meter}, 2.2$\times$ \textit{P4x4} for a single activation, and substituting the largest Flex-SFU configuration for ITA-Max raises this to 138$\times10^{3}$ \si{\micro\meter\squared}, 2.3$\times$ \textit{P4x4}. Rogue covers a comparable activation set, and more strategies, on one fabric at under half that area. Specialization allows better per-function efficiency, though less uniformly than expected. NACU is 12$\times$ smaller than Rogue and far more area efficient, yet Rogue is 2.3$\times$ faster on both functions and 1.2$\times$ more energy efficient at iso-technology. Among the reconfigurable designs, Flex-SFU leads Rogue on area efficiency, but Rogue matches its throughput range at a one-third lower frequency and without packed \ac{SIMD} for activations, which would add up to 4$\times$ on int8. Against PACE-Lite, Rogue's fully integer datapath yields up to 3.5$\times$ better energy efficiency at iso-technology.

In summary, while most accelerators either target a single activation or cover a broader set through a single approximation strategy, Rogue remains comparable to them in performance and efficiency while spanning \ac{PA}, \ac{PWLA}, \ac{PWPA} and base-substitution and supporting \ac{GeMM} on the same fabric.

\section{Conclusions}
We have presented Rogue, a streaming \ac{CGRA} whose \ac{ISA} spans the \ac{PA}, \ac{PWLA}, \ac{PWPA} and base-substitution families while supporting \ac{GeMM}. Tight \ac{DMA} integration keeps the system-level area overhead below 11.5\%, while Rogue reaches up to 137$\times$ speedup and 52$\times$ energy savings over an optimized RISC-V CPU baseline and runs a Transformer encoder layer 7.1$\times$ faster at 5.3$\times$ lower energy, confirming the benefit of a unified fabric. At iso-technology, it is also over 2$\times$ smaller and up to 5$\times$ more area efficient than the closest state-of-the-art \ac{CGRA}.

\section{Acknowledgments} 
This work was supported in part by the Swiss State Secretariat for Education, Research, and Innovation (SERI) through the SwissChips research project.

\bibliographystyle{IEEEtran}
\bibliography{bibliography}

\end{document}

%% file: acronyms.tex
\begin{acronym}
  \acro{ML}{Machine Learning}
  \acro{CGRA}{Coarse-Grain Reconfigurable Array}
  \acro{PEA}{Processing Element Array}
  \acro{DSMU}{Data Stream Management Unit}
  \acro{PE}{Processing Element}
  \acro{GeMM}{General Matrix Multiplication}
  \acro{ReLU}{Rectified Linear Unit}
  \acro{GELU}{Gaussian Error Linear Unit}
  \acro{Tanh}{Hyperbolic Tangent}
  \acro{SiLU}{Sigmoid Linear Unit}
  \acro{CNN}{Convolutional Neural Network}
  \acro{ViT}{Vision Transformer}
  \acro{MCU}{microcontroller unit}
  \acro{ISA}{Instruction Set Architecture}
  \acro{AI}{Artificial Intelligence}
  \acro{LSU}{Load-Store Unit}
  \acro{LSTM}{Long-Short Term Memory}
  \acro{INT}{integer}
  \acro{FP}{floating-point}
  \acro{DMA}{Direct Memory Access}
  \acro{LUT}{Look-Up Table}
  \acro{PA}{Polynomial Approximation}
  \acro{PWPA}{Piece-Wise Polynomial Approximation}
  \acro{PWLA}{Piece-Wise Linear Approximation}
  \acro{I/O}{Input/Output}
  \acro{FIFO}{First In First Out}
  \acro{DOO}{Delay-Operand Overlay}
  \acro{OpC}{Output-per-Cycle}
  \acro{MAE}{Mean Average Error}
  \acro{RMSE}{Root Mean Squared Error}
  \acro{SIMD}{Single Instruction Multiple Data}
  \acro{CGP}{Cartesian Genetic Programming}
  \acro{RF}{Register File}
  \acro{matmul}{matrix multiplication}
  \acro{II}{Initiation Interval}
  \acro{HPC}{High-Performance Computing}
  \acro{ALU}{Arithmetic Logic Unit}
  \acro{HWPE}{Hardware Processing Engine}
  \acro{LLM}{Large Language Model}
  \acro{GE}{Gate Equivalent}
  \acro{CSR}{Control and Status Register}
  \acro{DFG}{Data-Flow Graph}
\end{acronym}

%% file: tab/rogue_area.tex
\begin{table}[h]
  \centering
  \caption{Area breakdown of \acp{PE}, Rogue variants and related X\nobreakdashes-HEEP systems ($\times10^3$ \si{\micro\meter\squared}).}
  \label{tab:area}
  \resizebox{\linewidth}{!}{%
  \begin{threeparttable}
    \begin{tabular}{@{}l
                    S[table-format=3.1]
                    S[table-format=3.1]
                    S[table-format=3.1]
                    S[table-format=3.1]
                    S[table-format=3.1]
                    l@{}}
    \toprule
    & {\textit{S4x2}} & {\textit{P4x2}} & {\textit{S4x4}} & {\textit{P4x4}} & {X-HEEP} &  \\
    \midrule
    Conf. Regs.   & 5.4  & 5.3  & 10.7 & 10.7 & \multicolumn{1}{!{\vrule\vrule}c}{9.6}   & CV32E40P \\
    \ac{DSMU}     & 1.1  & 1.1  & 2.2  & 2.2  & \multicolumn{1}{!{\vrule\vrule}c}{375.9} & Memory \\
    PEA           & 16.4 & 23.0 & 34.2 & 47.0 &
      \multicolumn{1}{!{\vrule\vrule}c}{\makecell{5.0 (2 Ch.) \\ 10.0 (4 Ch.)}} & DMA \\
                  &      &      &      &      & \multicolumn{1}{!{\vrule\vrule}c}{47.5}  & Other Periph. \\
                  &      &      &      &      & \multicolumn{1}{!{\vrule\vrule}c}{79.2}  & Pad Ring \\
    \midrule
    Total Area
      & \multicolumn{1}{c}{\makecell{22.9 \\ (+4.4\%)}}
      & \multicolumn{1}{c}{\makecell{29.4 \\ (+5.7\%)}}
      & \multicolumn{1}{c}{\makecell{47.1 \\ (+9.0\%)}}
      & \multicolumn{1}{c}{\makecell{59.9 \\ (+11.5\%)}}
      & \multicolumn{1}{!{\vrule\vrule}c}{\makecell{517.2 (2 Ch.) \\ 522.2 (4 Ch.)}}
      & Total Area \\
    \midrule\midrule
    \multicolumn{7}{@{}l}{\textit{Rogue PEs Extensions}} \\
    \midrule
    {Base} & {SIMD Add} & {SIMD Mul} & {Pipe. Div.} & {Ser. Div.} & {Breakpoint} & {Coefficient} \\
    \midrule
    {1.77} & {2.04}     & {1.88}     & {5.85}       & {2.60}      & {1.79}       & {1.93}        \\
     & {1.15$\times$}    & {1.06$\times$}    & {3.31$\times$}      & {1.47$\times$}     & {1.01$\times$}      & {1.09$\times$}       \\
    \bottomrule
    \end{tabular}
  \end{threeparttable}}
\end{table}

%% file: tab/comparison_sota.tex
\begin{table*}[t]
    \centering
    \caption{Comparison with State-of-the-Art Accelerators.}
    \label{tab:comparison-sota}
    \resizebox{0.97\textwidth}{!}{%
    \begin{threeparttable}
        \begin{tabular}{@{}cccccccccccc@{}}
            \toprule
            Work  &
            \makecell[c]{Precision} &
            \makecell[c]{Activations\tnote{f}} &
            \makecell[c]{Linear} &
            \makecell[c]{Computation} &
            \makecell[c]{Node\\{[}\si{\nano\meter}{]}} &
            \makecell[c]{Frequency\\{[}\si{\mega\hertz}{]}} &
            \makecell[c]{Area\tnote{g}\\{[}$\times10^3$\si{\micro\meter\squared}{]}}  &
            \makecell[c]{Power\\{[}\si{\milli\watt}{]}}  &
            \makecell[c]{Throughput\\{[}GEval/s{]}} &
            \makecell[c]{Energy\\Efficiency\tnote{g}\\{[}GEval/s/W{]}}  &
            \makecell[c]{Area\\Efficiency\tnote{g}\\{[}GEval/s/$mm^{2}${]}}  \\
            \midrule

            \makecell[c]{ITA~\cite{islamoglu_ita_2023}\tnote{a}} &   
            INT8	&                                       
            IM &                                            
            \ding{51} &
            \makecell[c]{Base conv.} &                      
            22 &                                            
            500 &                                           
            \makecell[c]{173 \textit{131}\\(6) \textit{4.6}} &  
            \makecell[c]{60.5\\(0.85)} &                    
            0.5 &                                  
            \makecell[c]{588.2\\\textit{1013}} &                
            \makecell[c]{83.3\\\textit{110}} \\           

            \midrule

                        NACU~\cite{baccelli_nacu_2020} &                  
            FxP16 &                                         
            S, T &                                           
            \ding{55} &
            PWLA &                                          
            28 &                                            
            267 &                                           
            \makecell[c]{9.7\\\textit{4.9}} &            
            \makecell[c]{2.2-2.0} &              
            0.3-0.3 &                                    
            \makecell[c]{125.0-138.5\\\textit{356-395}} &   
            \makecell[c]{270-270\\\textit{529}} \\       

            \midrule

            Flex-SFU~\cite{andri_flex-sfu_2025}\tnote{b} &                       
            INT,FP8/16/32 &                                 
            \makecell[c]{Sm, G, T,\\S, SL, HS} &            
            \ding{55} &
            PWPA &                                          
            28 &                                            
            500 &                                           
            \makecell[c]{4.0-22.9\\\textit{2.0-11.7}} &   
            0.4-1.6 &                        
            0.5-2 &                                         
            \makecell[c]{374.6-4500\\\textit{1068-12830}} &     
            \makecell[c]{1176-1250\\\textit{2306-2451}} \\ 

            \midrule

            PACE-Lite~\cite{prasad_pace-lite_2025} &        
            FP16/32, BFP32 &                                
            Sm, G &                                     
            \ding{55} &
            PWPA &                                          
            12 &                                            
            1000 &                                          
            \makecell[c]{28.1\\\textit{39.6}} &          
            55.7 &                                          
            15.6 &                                          
            \makecell[c]{280\\\textit{200}} &                   
            \makecell[c]{557.1\\\textit{395}} \\         

\midrule

            \makecell[c]{NX-CGRA~\cite{prasad_nx-cgra_2025}\tnote{c}\\$\langle G/Sm//MM\rangle$} &   
            INT8/16/32 &                                    
            G, Sm &                                     
            \ding{51} &
            - &                                             
            22 &                                            
            200 &                                           
            \makecell[c]{178\\\textit{135}} &             
            \makecell[c]{1.6/1.6//1.5} &                       
            \makecell[c]{0.6/1.1//3.0\\ GOPS} &             
            \makecell[c]{390/680//2010\\\textit{672/1171//3462}\\ GOPS/W} &  
            \makecell[c]{3.6/6.2//17.1\\\textit{4.7/8.2//22.5}\\ GOPS/$mm^{2}$} \\

            \midrule

            \makecell[c]{PICACHU~\cite{qin_picachu_2025}\tnote{d}} &  
            INT16/32, FP16/32 &                             
            Sm, G &                        
            \ding{55} &
            \ac{PA} &               
            45 &                                            
            1000 &                                          
            \makecell[c]{1000\\\textit{190}} &           
            64.2 &                                          
            - &                                             
            - &                                             
            - \\                                            

            \midrule

            \makecell[c]{\textbf{Rogue}\tnote{e}\\$\langle G/Sm/S/T/IM//MM\rangle$} &                        
            INT8/16/32 &                                        
            \makecell[c]{Sm, G, T,\\S, SL, R, IM} &    
            \ding{51} &
            \makecell[c]{\ac{PA}/\ac{PWLA}/\ac{PWPA}\\Base conv.} &                                  
            16  &                                                
            330 &                                               
            60  &                                             
            \makecell[c]{1.0/1.1/1.5/\\1.6/1.2//1.0} &   

            \makecell[c]{0.7/0.2/0.7/\\0.7/0.3//6.7} &   

            \makecell[c]{700/182/467/\\437/250//6700} &   

            \makecell[c]{11.6/3.3/11.6/\\11.6/5.0//111.7} \\   


            \bottomrule
        \end{tabular}

        \begin{tablenotes}
            \scriptsize
            \item[a] \textit{ITA}: area/power for the whole ITA, in parentheses for ITA-Max alone, to which throughput and efficiency refer.
            \item[b] \textit{Flex-SFU}: excludes the host vector processing unit it is integrated into.
            \item[c] \textit{NX-CGRA}: results follow the $\langle\rangle$ order, throughput as GOPS.
            \item[d] \textit{PICACHU}: \ac{CGRA} only, plugged into a separate systolic array for \ac{GeMM}. Also supports GeGLU, SwiGLU, LayerNorm and RMSNorm normalization and RoPE embedding. No per-evaluation throughput reported.
            \item[e] \textit{Rogue}: results follow the $\langle\rangle$ order.
            \item[f] \ac{Tanh} (T), Sigmoid (S), Softmax (Sm), \ac{GELU} (G), \ac{ReLU} (R), \ac{SiLU} (SL), HardSwish (HS), ITA-Max (IM); for each accelerator, the activations evaluated in its paper are listed. MM is \ac{matmul} in the $\langle\rangle$ order.
            \item[g] Italic values are scaled to 16 \si{\nano\meter} following~\cite{stillmaker_scaling_2017}. Power and throughput are at native nodes.
        \end{tablenotes}
    \end{threeparttable}}
\end{table*}